\documentclass[aps,prl,showpacs,twocolumn,preprintnumbers,amsmath,amsfonts,bm]{revtex4}
\usepackage[dvips]{graphicx}
\usepackage{dcolumn}

\begin{document}

\title{Damage Spreading and Criticality in Finite Random Dynamical Networks}

\author{Thimo Rohlf$^{1}$, Natali Gulbahce$^{2}$ and Christof Teuscher$^{3}$}

\affiliation{
$^1$Santa Fe Institute, 1399 Hyde Park Road, Santa Fe, NM 87501, U.S.A \\
$^2$Los Alamos National Laboratory, Theoretical Division and Center for Nonlinear 
Studies, MS B284, Los Alamos, NM 87545, U.S.A.\\
$^3$Los Alamos National Laboratory, CCS-3, MS B287, Los Alamos, NM 87545, U.S.A.
}
\date{\today}

\begin{abstract}
We systematically study and compare damage spreading at the sparse
percolation (SP) limit for random boolean and threshold networks with
perturbations that are independent of the network size $N$. This limit
is relevant to information and damage propagation in many
technological and natural networks. Using finite size scaling, we
identify a new characteristic connectivity $K_s$,
at which the average number of damaged nodes $\bar d$, after a large number
of dynamical updates, is independent of
$N$.  Based on marginal damage spreading, we determine the critical
connectivity $K_c^{sparse}(N)$ for finite $N$ at the SP limit and show
that it systematically deviates from $K_c$, established by the
annealed approximation, even for large system sizes.  Our findings can
potentially explain the results recently obtained for
gene regulatory networks and have important implications for the
evolution of dynamical networks that solve specific computational or
functional tasks.
\end{abstract}


\maketitle

Random boolean networks (RBN) were originally introduced as simplified
models of gene regulation \cite{Kauffman69,Kauffman93}, focusing on a
system-wide perspective rather than on the often unknown details of
regulatory interactions \cite{Bornholdt2001}. In the thermodynamic
limit, these disordered dynamical systems exhibit a dynamical
order-disorder transition at a sparse critical connectivity $K_c$
\cite{DerridaP86}; similar observations were made for sparsely
connected random threshold (neural) networks (RTN)
\cite{Kuerten88a,Kuerten88b,RohlfBornhol02}.  For a finite system size
$N$, the dynamics of both systems converge to periodic attractors
after a finite number of updates. At $K_c$, the phase space structure in
terms of attractor periods \cite{AlbertBaraBoolper00}, the number of
different attractors \cite{SamuelsonTroein03} and the distribution of
basins of attraction \cite{Bastola98} is complex, showing many
properties reminiscent of biological networks \cite{Kauffman93}.

Often, one is interested in the response of dynamical networks to
external perturbations; because these signals can disrupt the generic
dynamical state (fixed point or periodic attractor) of the network,
they are usually referred to as ``damage.'' This type of study has
numerous applications, e.g., the spreading of disease through a
population \cite{Pastor2001,Newman2002}, the spreading of a computer
virus on the internet \cite{Cohen2003}, failure propagation in power
grids \cite{Sachtjen2000}, or the perturbation of gene expression
patterns in a cell due to mutations \cite{RamoeKesseliYli06}.
Mean-field approaches, e.g., the annealed approximation (AA)
introduced by
Derrida and Pomeau \cite{DerridaP86}, allow for an analytical treatment of
damage spreading and exact determination of the critical connectivity $K_c$
under various constraints \cite{SoleLuque95,LuqueSole96}.  It has been shown
that local, mean-field-like rewiring rules coupled to order parameters of the
dynamics can drive both RBN and RTN to self-organized criticality
\cite{BornholRohlf00,BornholRoehl2003,LiuBassler2006}.

Mean-field approximations of RBN/RTN dynamics rely on the assumption that
$N\to\infty$ and study the rescaled damage $d(t)/N$ (where $d(t)$ is the
number of damaged nodes at time $t$). For an application to real-world problems,
these limits are often not very relevant.  Going beyond the framework of
AA, a number of recent studies focus on the finite-size scaling of
(un-)frozen and/or relevant nodes in RBN with respect to $N$ 
\cite{KaufmanMihaljevDrossel05,Mihaljev2006}; only few studies, however, 
consider finite-size scaling of damage spreading in RBN
\cite{RamoeKesseliYli06,SamuelsonSocolar06}. Here, of particular interest is
the ``sparse percolation (SP) limit'' \cite{SamuelsonSocolar06}, where the initial
perturbation size $d(0)$ does {\em not} scale up with network size $N$, i.e.,
the relative size of perturbations tends to zero for large $N$. This limit
applies to many of the above-mentioned real-world networks (e.g., the spread of
a new computer virus on the internet launched from a single computer). In this
letter, we systematically study finite-size scaling of damage spreading in the
SP limit for both RBN and RTN. We identify a new characteristic point $K_s$, 
where the expectation value of the number of damaged nodes after large number
of dynamical updates is independent of $N$.
By the definition of marginal damage spreading, we
introduce a new approach to estimate the critical connectivity $K_c(N)$ for
finite $N$, and present
evidence that, even in the large $N$ limit, the critical connectivity for SP
systematically deviates from the predictions of mean-field theory.

First, let us define the dynamics of RBN and RTN.
A RBN is a discrete dynamical system composed of $N$ automata. 
Each automaton is a Boolean variable with two possible states: 
$\{0,1\}$, and the dynamics is such that
\begin{equation}
{\bf F}:\{0,1\}^N\mapsto \{0,1\}^N, 
\label{globalmap}
\end{equation} 
where ${\bf F}=(f_1,...,f_i,...,f_N)$, 
and each $f_i$ is represented by a look-up table of $K_i$ inputs
randomly chosen from the set of $N$ automata. Initially, $K_i$ 
neighbors and
a look-table are assigned to each automaton at random.

An automaton state $ \sigma_i^t \in \{0,1\}$ is updated using
its corresponding Boolean function:
\begin{equation}
\sigma_i^{t+1} = f_i(x_{i_1}^t,x_{i_2}^t, ... ,x_{i_{K_i}}^t).
\label{update}
\end{equation}
We randomly initialize the states of the automata (initial 
condition of the RBN). The automata are updated synchronously using their
corresponding Boolean functions.
%
%
The second type of discrete dynamical system we study is RTN. An RTN consists 
of $N$ randomly interconnected binary sites (spins) with states $\sigma_i=\pm1$.
For each site $i$, its state at time $t+1$ is a function of the inputs it receives 
from other spins at time $t$:
\begin{eqnarray} 
\sigma_i(t+1) = \mbox{sgn}\left(f_i(t)\right) 
\end{eqnarray}  
with 
\begin{eqnarray} 
f_i(t) = \sum_{j=1}^N c_{ij}\sigma_j(t) + h.  
\end{eqnarray}
The $N$ network sites are updated synchronously. In the following
discussion the threshold parameter $h$ is set to zero. The interaction weights
$c_{ij}$ take discrete values $c_{ij} = +1$ or $-1$ with equal
probability. If $i$ does not receive signals from $j$, one has $c_{ij} = 0$.\\

\begin{figure}[htb]
\begin{center}
\resizebox{85mm}{!}{\includegraphics{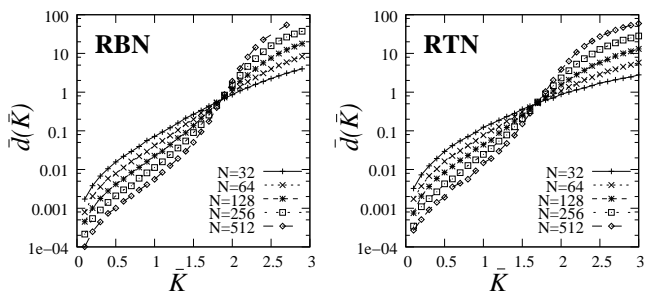}}
\end{center}
\caption{\small Average Hamming distance (damage) $\bar{d}$ after 200 system
updates, averaged over 10000 randomly generated networks for each value of
$\bar{K}$, with 100 different random initial conditions and one-bit perturbed
neighbor configurations for each network. For both RBN and RTN, all curves for
different $N$ approximately intersect in a characteristic point $K_s$.}
\label{dofkfig} 
\end{figure}

\begin{figure}[htb]
\begin{center}
\resizebox{85mm}{!}{\includegraphics{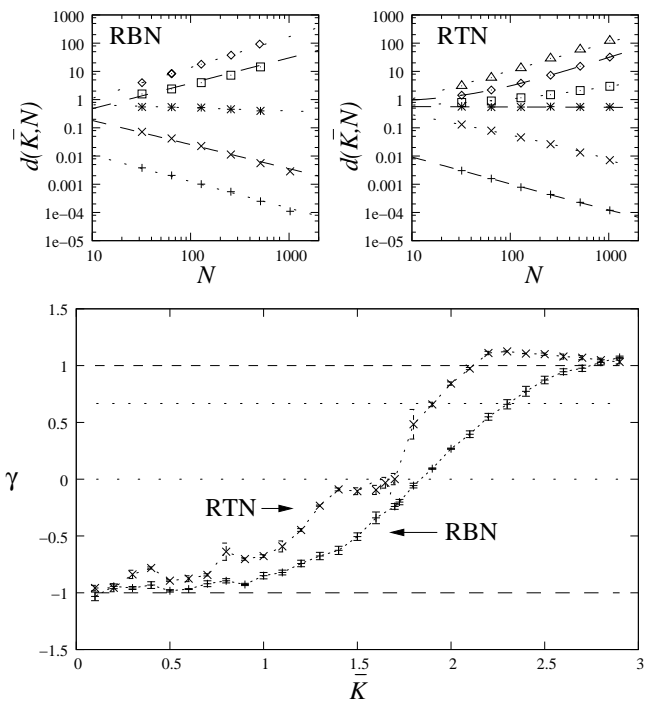}}
\end{center}
\caption{\small {\em Upper panels:} $\bar{d}$ as a function of $N$ for 
different $\bar{K}$: $\bar{K}= 1.0$ (+), $\bar{K}=
1.5$ (x), $\bar{K}= 1.8$ (RBN, *) and $\bar{K}= 1.7$ (RTN,
*), $\bar{K}= 2.1$ ($\Box$, RBN) and $\bar{K}= 1.9$  ($\Box$, RTN), $\bar{K}= 2.4$ ($\triangle$, RTN) and $\bar{K}= 2.6$ ($\diamond$).
The lines are fits of Eq. \ref{fs_eqn} to the data. {\em Lower panel}: Scaling
exponents $\gamma(\bar{K})$ as a function of $\bar{K}$, as
obtained from fits of Eq. \ref{fs_eqn} for RBN (+) and RTN (x). The
dashed/dotted lines mark the asymptotes as discussed in the text. }
\label{gammaofkfig}
\end{figure}

\begin{figure}[htb]
\begin{center}
\resizebox{85mm}{!}{\includegraphics{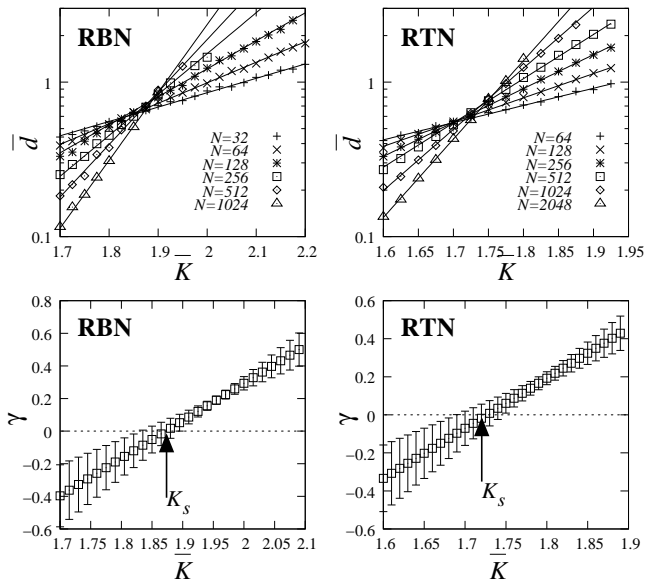}}
\end{center}
\caption{\small{\em Upper panels:} $\bar{d}(\bar{K})$ in semi-log
  scale, as obtained from high-precision simulations near $K_c$
  (ensemble size: 50000 random networks with 100 random initial
  conditions for each data-point, transient time: 5000 updates).
  Lines are fits of Eq.  \ref{exp_eqn}. {\em Lower
    panels:} Scaling exponents $\gamma$ derived from equating
  Eq. \ref{fs_eqn} and \ref{exp_eqn}, with corresponding
  errorbars. The intersection with $\gamma = 0$ (dashed line) defines
  $K_s$.} \label{ksfinebothfig}
\end{figure}

\begin{figure}[htb]
\begin{center}
\resizebox{85mm}{!}{\includegraphics{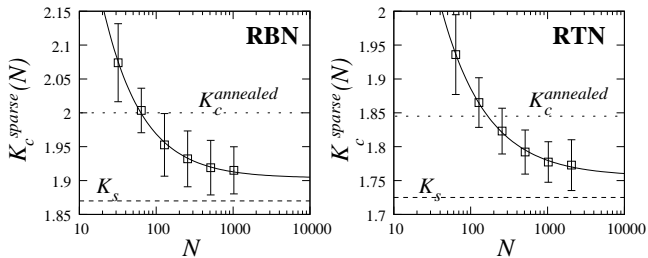}}
\end{center}
\caption{\small The critical connectivity $K_c^{sparse}(N)$ in the SP limit as
a function of $N$, calculated from Eq. \ref{kc_eqn}.
Curves are
power-law fits according to Eq. \ref{kcscale_eqn}, straight dashed lines mark $K_c^{annealed}$ and $K_s$ for
RBN and RTN, respectively.} 
\label{kcofNfig}
\end{figure}

{\em Results.} We first study the expectation value $\bar{d}$ of damage,
quantified by the Hamming distance of two different system configurations,
after a large number $T$ of system updates.  Let $\mathcal{N}$ be a randomly
sampled set (ensemble) of $z_N$ networks with average degree $\bar{K}$, 
$\mathcal{I}_n$ a set of $z_I$ random initial conditions tested 
on network $n$, and $\mathcal{I^{\prime}}_n$ a set of $z_I$ random initial 
conditions differing in one randomly chosen bit from these initial conditions. 
Then we have
\begin{equation}
 \bar{d} = \frac{1}{z_N\,z_I}\sum_{\stackrel{n=1}{\mathcal{N}_n \in \mathcal{N}}}^{z_N}\sum_{\stackrel{i=1}{\vec{\sigma}_i \in \mathcal{I}_n, \vec{\sigma}^{\prime}_i \in \mathcal{I}^{\prime}_n} }^{z_I} d_i^n(T),\label{d_aver_eq}
\end{equation}
where $d_i^n(T)$ is the measured Hamming distance after $T$ system updates.
Fig. \ref{dofkfig} shows $\bar{d}$ as a function of the average connectivity
$\bar{K}$ for different network sizes $N$ by using a random ensemble
for statistics.  For both RBN and RTN, the observed functional behavior strongly
suggests that the curves approximately intersect at a common point $(K_s, d_s)$,
where the observed Hamming distance for large $t$ is independent of the system
size $N$. 

To verify this finding, let us now study the finite size scaling behavior of
$\bar{d}$ in this (SP) limit. For $\bar{K} \to 0$ and for large
$\bar{K}$, it is straightforward to estimate the asymptotic scaling.
In the case $\bar{K} \to 0$, non-zero damage can only emerge if the
initial perturbation 
hits a short loop of oscillating nodes (most likely a self-connection or a 
loop of length two, longer loops can be neglected). The {\it a priori} 
probability to generate these loops is $\sim 1/N^2$, and their number is 
proportional to the total number of links, $\bar{K} N$. Hence, we 
expect $\bar{d} \sim \bar{K} N/N^2 \propto
1/N$. For large $\bar{K}$, damage percolates through the system,
consequently, avalanche sizes are bounded only by the size of the system, and we
expect $\bar{d} \sim N$. At criticality, the frozen core of the network always
remains undamaged; for this reason, $\bar{d}$ should be limited by the number
of {\em unfrozen} nodes $n_u$. Recent studies predict $n_u \sim N^{2/3}$ at
$K_c$ for Boolean networks \cite{KaufmanMihaljevDrossel05}, hence, we expect
$\bar{d} \le N^{2/3}$ at $K_c$.  For arbitrary $\bar{K}$, we make 
the scaling ansatz
 
\begin{equation}
\bar{d}(\bar{K}, N) = a(\bar{K})\cdot N^{\gamma(\bar{K})} + d_0(\bar{K}),\,-1 \le \gamma \lesssim 1. \label{fs_eqn}
\end{equation}

The two upper panels of Fig. \ref{gammaofkfig} illustrate that for both RBN
and RTN, the numerically measured finite size scaling for $0 < \bar{K}
\le 3$ obeys this scaling ansatz very well. In both cases, at some $K_s$
slightly below $K_c$, we find a transition of $\gamma$ from negative to positive
values (Fig. \ref{gammaofkfig}, lower panel). The exact determination of the
point $(K_s, d_s)$ where $\gamma \approx 0$ is difficult; because of the slow
emergence of large damage events near $K_c$, measurements with finite $T$ can
substantially underestimate $\bar{d}$ (in particular for $N \ge 512$). We
performed high precision numerical experiments in the interval $1.6 \le \bar{K} \le 2.1$, waiting for $T=5000$ update time steps to let the network
dynamics relax after the initial one-bit perturbation; these simulations
conclusively show an exponential dependence $\bar{d} \propto
\exp(c(N)\cdot\bar{K})$ in this interval, with a constant $c(N)$
depending only on $N$ (Fig. \ref{ksfinebothfig}, upper panels). This
exponential dependence becomes apparent with the following
assumptions: an increase $\Delta\bar{d}$ of the average damage is proportional
to $\bar{d}$ itself (damage can generate new damage), to an increase
$\Delta\bar{K}$ of the average connectivity, and to some function of the system
size $N$. Actually, it cannot be directly proportional to $N$, because
nodes that are part of the frozen core always remain undamaged asymptotically; hence, a rough upper
limit is given by the number of nonfrozen nodes,  which at $K_c$ scales as $N^{2/3}$.
A lower bound can be derived by the number of relevant nodes, that almost certainly
propagate damage, i.e. $N^{1/3}$ \cite{KaufmanMihaljevDrossel05}.
To summarize, we approximate 
\begin{equation} \Delta\bar{d}
\approx c(N)\,N^{\alpha}\,\bar{d}\,\,\Delta\bar{K}, 
\end{equation}
with $1/3 \lesssim \alpha \lesssim 2/3$;
replacing $\Delta\bar{d}$ and $\Delta\bar{K}$ with differentials and
integrating yields
\begin{equation}
\bar{d}(\bar{K}, N) \approx c_1(N)\,\exp{[c_2(N)\,N^{\alpha}\,\bar{K}]}. \label{exp_eqn}
\end{equation}
In simulations, we find $\alpha \approx 0.42$, which is well within the range we expect from our theoretical considerations as discussed above.

We now apply this dependence to obtain high-accuracy fits of Eq. \ref{fs_eqn} in the interval $1.6 \le \bar{K} \le 2.1$ (Fig. \ref{ksfinebothfig}, lower panels); these fits yield
\begin{equation}
(K_s^{RBN}, d_s^{RBN}) = (1.87 \pm 0.04, 0.65 \pm 0.05)
\end{equation}
for RBN and, correspondingly,
\begin{equation}
(K_s^{RTN}, d_s^{RTN}) = (1.725 \pm 0.035, 0.52 \pm 0.04)
\end{equation}
for RTN.

Interestingly, $K_s$ is close to, but distinct from the critical
connectivities $K_c^{RBN} =2$ and $K_c^{RTN} =1.845$, as predicted by
mean-field theory. However, a natural comparison has to consider
possible deviations of $K_C$ at the SP limit from these values.  An
intuitive definition of criticality for finite $N$ can be formulated
in terms of {\em marginal damage spreading}. If at time $t$ one bit is
flipped, one requires at time $t+1$ \cite{LuqueSole96,RohlfBornhol02}
\begin{equation}
\bar{d}(t+1) = \langle p_s\rangle(K_c) K_c = 1, 
\end{equation}
where $\langle p_s\rangle(\bar{K})$ is the average damage propagation
probability. Naturally, the iteration of this map implies $\bar{d} =
1$ for all $t$.  Note that the relation: $\langle p_s\rangle(K_c) K_c = 1$
is exact only in the framework of the AA. In the SP limit, we instead
have to set the right hand side of Eq. \ref{exp_eqn} to unity;
inversion then leads to \begin{equation} K_c^{sparse}(N) \approx
  -\frac{\ln{c_1(N)}}{c_2(N)N^{\alpha}}.
\label{kc_eqn} 
\end{equation}
Fig. \ref{kcofNfig} shows $K_c^{sparse}(N)$, using the values $c_1(N), c_2(N)$
obtained from numerical fits of Eq. \ref{exp_eqn} for both RBN and RTN. We find
that both systems, in a very good approximation, obey the scaling relationship
\begin{equation}
K_c^{sparse}(N) \approx b\cdot N^{-\delta} + K_c^{\infty}\label{kcscale_eqn}
\end{equation}
with $b = 3.27 \pm 0.79$, $\delta = 0.85 \pm 0.07$ and $K_c^{\infty} = 1.90375 \pm 0.005$ for RBN and $b = 3.853 \pm 0.76$,
$\delta = 0.736 \pm 0.05$ and $K_c^{\infty} = 1.75598 \pm 0.005$ for RTN.
Hence, in the limit $N \to\infty$, we have 
\begin{equation} K_c^{\infty, RBN} =
1.90375 \pm 0.005
\end{equation}
for RBN, and for RTN
 \begin{equation}
 K_c^{\infty, RTN} = 1.75598 \pm 0.005.
\end{equation}
Thus for both RBN and RTN, in the SP limit and for
$N\to\infty$, the critical connectivity is considerably below the value $K_c$
predicted by the annealed approximation (notice that for {\em small} $N < 128$,
however, $K_c^{sparse}(N) > K_c^{annealed}$). 

It is beyond the scope of this letter to discuss possible causes for
these deviations in detail (this will be accomplished in a long
paper). In simulations, however, we find that the statistical
distributions of damage sizes in the SP limit are highly skewed, with
most configurations leading to vanishing damage, and a fat tail of
large damage events. These skewed distributions imply that with finite
sampling, we always underestimate $\bar{d}$, and hence the true
$K_c^{sparse}(N)$ and $K_s$ will deviate even stronger from the
AA. Also, local fluctuations in damage propagation cannot be neglected
in this limit, as it is assumed in mean-field approaches.

{\em Discussion.} 
We investigated finite size scaling of damage spreading in both RBN and RTN
near the sparse percolation (SP) transition. We find that the average damage
$\bar{d}$, quantified in terms of the Hamming distance of initially nearby
system states, scales $\propto N^{\gamma(\bar{K})}$ over the whole
range of sparse connectivities $0 < \bar{K} \le 3$ studied in this
letter. The scaling exponents $\gamma$ show a cross-over from negative to
positive values at characteristic points $K_s^{RBN}$ and $K_s^{RTN}$ 
below the critical points $K_c^{RBN}$ and $K_c^{RTN}$. We estimated the
critical connectivities $K_c^{sparse}(N)$ based on marginal damage spreading,
and found deviations from the annealed approximation even in the limit of
large $N$. Interestingly, recent studies suggest that gene regulatory networks
appear to be in the ordered regime and reside slightly below the phase
transition between order and chaos \cite{RamoeKesseliYli06}, while theory had
proposed the critical line to be an evolutionary attractor
\cite{Kauffman69,Kauffman93}.  Our study may contribute a possible explanation
to these observations: in finite networks, scaling of damage with increasing
$N$ (e.g., as a consequence of gene duplications) can be expected to be under
strict selective control. Hence, the need for robust systems might drive
evolution to $K_s$ rather than to $K_c$. At the same time, however, $K_s$ is
close enough to criticality to enable rich dynamical behavior, as required by
biological cells. 
Certainly this question of where in the dynamical phase space biological
networks reside cannot be answered conclusively, given the current state of
systems biology. Experimental studies of regulatory networks and further 
studies of in-silico evolutionary processes that adapt dynamical 
networks, e.g., RBNs or RTNs, to robustly solve specific computational and 
functional tasks are required.

{\em Acknowledgements.}  This work was partly carried out
under the auspices of the NNSA of the U.S. DOE at LANL under Contract
No. DE-AC52-06NA25396.

\end{document}